\documentclass[5p]{elsarticle}

\usepackage{lineno,hyperref}
\modulolinenumbers[5]

\journal{Computational Materials Science}
\usepackage{epsfig}
\usepackage{amsmath}
\usepackage{tcolorbox}
\usepackage{booktabs}

\usepackage{color}

\newcommand{\cpp}{{C\nolinebreak[4]\hspace{-.05em}\raisebox{.4ex}{\tiny\bf ++}}}

\begin{document}

\begin{frontmatter}

\title{Rapid multiphase-field model development using a modular free energy based approach with automatic differentiation in MOOSE/MARMOT}

\author[inl1]{D.~Schwen\corref{correspondingauthor}}
\ead{daniel.schwen@inl.gov}
\cortext[correspondingauthor]{Corresponding author}

\author[inl1]{L.~K.~Aagesen}
\author[inl2]{J.~W.~Peterson}
\author[inl1,penn]{M.~R.~Tonks}

\address[inl1]{Fuels Modeling and Simulation Department, Idaho National Laboratory, Idaho Falls, ID 83415, United States}
\address[penn]{Mechanical and Nuclear Engineering Department, Pennsylvania State University, University Park, PA 16802, USA}
\address[inl2]{Modeling and Simulation Department, Idaho National Laboratory, P.O. Box 1625, Idaho Falls, ID 83415, USA} 

\begin{abstract}
We present a novel phase-field model development capability in the
open source MOOSE finite element framework.  This facility is based on
the ``modular free energy'' approach in which the phase-field
equations are implemented in a general form that is logically
separated from model-specific data such as the thermodynamic free
energy density and mobility functions.  Free energy terms contributing
to a phase-field model are abstracted into self-contained objects that
can be dynamically combined at simulation run time. Combining multiple chemical and mechanical free energy
contributions expedites the construction of coupled phase-field,
mechanics, and multiphase models. This approach
allows computational material scientists to focus on implementing new
material models, and to reuse existing solution algorithms and data
processing routines. A key new aspect of the rapid phase-field
development approach that we discuss in detail is the automatic
symbolic differentiation capability. Automatic symbolic
differentiation is used to compute derivatives of the free energy
density functionals, and removes potential sources of human error
while guaranteeing that the nonlinear system Jacobians are accurately
approximated.  Through just-in-time compilation,
we greatly reduce the computational expense of evaluating the
differentiated expressions. The new capability is demonstrated for a
variety of representative applications.
\end{abstract}

\begin{keyword}
  phase-field \sep finite element \sep automatic differentiation
  \PACS \sep 46.15.-x \sep 05.10.-a \sep 02.70.Dh
  \MSC[2010] 65-04 \sep 65Z05
\end{keyword}

\end{frontmatter}
%

\section{Introduction}
The phase-field method is a well-established tool for simulating the coevolution of microstructure and physical properties at the mesoscale~\cite{chen_phase-field_2002, moelans2008introduction}. In the phase-field method, the microstructure is described by a system of continuous variables, also called order parameters.  Microstructure interfaces are approximated using a finite width, and the order parameters vary smoothly over the interfaces.
In isolated systems, the evolution of these variables leads to a monotonically decreasing free energy as a function of time.
The phase-field method has been used to model a large range of physical phenomena, including solidification~\cite{warren1995prediction, karma1996phase}, phase transformation~\cite{wheeler1992phase, kim_phase-field_1999}, and grain growth~\cite{fan1997diffusion, moelans2008quantitative}.

The microstructure evolution is governed by partial differential equations (PDEs) that depend on the derivatives of a functional defining the free energy of the system in terms of the phase-field variables. The phase-field PDEs have been solved using the finite difference~\cite{wheeler1993computation}, finite volume~\cite{guyer2009}, and finite element (FEM) methods~\cite{takaki2005phase, tonks_object-oriented_2012}, as well as spectral methods based on the fast Fourier transform~\cite{chen1998applications}. The finite difference method is the easiest to implement, while the spectral methods are computationally efficient for small systems and offer better convergence properties. The finite volume and finite element methods are the most flexible, allowing a large range of boundary conditions, domain shapes, and coupling to other physics.

No matter which method is used to solve the phase-field equations, their basic form remains the same for most models: only the free energy functional
and the mobility expressions change when modeling different materials and physical phenomena.
In this work, we continue the efforts described in~\cite{tonks_object-oriented_2012} where a free energy based approach to phase-field model development is employed.
This strategy takes advantage of the ``fixed'' form of the governing phase-field equations, and restricts the task of programming new phase-field models to the
task of programming new free energy functionals. To the authors' knowledge, this specific approach is not currently followed by other open source phase-field
software development efforts~\cite{Puchala2016,logg2012automated,guyer2009,mmsp}, however it is also a fairly natural one that could be implemented retroactively
in existing code bases.

The free energy based approach employs three capabilities which simplify and accelerate the development of new multiphase-field models: (i) a symbolic differentiation module which eliminates the need to compute derivatives (up to third-order) of the free energy, (ii) a system to modularize, recombine, and reuse the various free energy contributions across models, and (iii) a generic framework for multiphase-field simulations in which new free energy modules can be employed. The sum of these components represents an important step toward a modular and general phase-field approach.

\section{Phase-Field method summary}
In the phase-field method, the evolution of non-conserved order parameters $\eta_j$ (e.g.\ phase regions and grains)
is governed by the Allen--Cahn~\cite{AllenCahn} equation~\eqref{eq:ac} and the evolution of conserved order
parameters $c_i$ (e.g.\ concentrations) is governed by the Cahn--Hilliard~\cite{CahnHilliard} equation~\eqref{eq:ch}:
\begin{align}
  \label{eq:ac}
  \frac{\partial \eta_j}{\partial t} &= - L_j \frac{\delta F}{\delta \eta_j}, & j=1,\ldots,N_{\eta}
  \\
  \label{eq:ch}
  \frac{\partial c_i}{\partial t} &= \nabla \cdot \left(M_i \nabla \frac{\delta F}{\delta c_i}\right), & i=1,\dots,N_c
\end{align}
Here, $F$ is the total free energy of the system, which can be formulated as a volume integral
\begin{equation}
  \label{eq:F}
  F = \int_\Omega \left( f_\text{loc} + f_\text{gr} + E_\text{d} \right) \, \text{d}V,
\end{equation}
where $\Omega$ is the simulation domain,
\begin{align}
  \label{eq:floc_general}
  f_\text{loc} \equiv f_\text{loc}(\eta_1,\eta_2,\ldots, c_1,c_2,\ldots)
\end{align}
is the local free energy density, and
\begin{align}
  f_\text{gr} \equiv f_\text{gr}(\nabla\eta_1,\nabla\eta_2,\ldots, \nabla c_1,\nabla c_2,\ldots)
\end{align}
is the gradient energy contribution.
$E_\text{d}$ is the free energy contribution due to external driving forces.
In general, the total free energy depends on \emph{all} of the conserved and non-conserved order parameters and their gradients.
Computing the variational derivatives in~\eqref{eq:ac} and~\eqref{eq:ch} yields terms which depend on the derivatives of the local free energy density, $f_\text{loc}$, with respect to all order parameters, i.e.
\begin{align}
  \label{eq:ac_strong}
  \frac{\partial \eta_j}{\partial t} &=
  - L_j \left( \frac{\partial f_\text{loc}}{\partial \eta_j}
  - \nabla\cdot \frac{\partial f_\text{gr}}{\partial\nabla \eta_j }
  + \frac{\partial E_{d}}{\partial \eta_j} \right)
  \\
  \label{eq:ch_strong}
  \frac{\partial c_i}{\partial t} &=
  \nabla \cdot M_i \nabla \left( \frac{\partial f_\text{loc}}{\partial c_i}
  -\nabla\cdot \frac{\partial f_\text{gr}}{\partial\nabla c_i}
  + \frac{\partial E_{d}}{\partial c_i} \right).
\end{align}
Note that the thermodynamics of the modeled system are determined by the local free energy density $f_\text{loc}$, while the gradient contribution $f_\text{gr}$ produces diffuse-width interfaces and contributes only to the interfacial energy. The gradient contribution is often a known functional that exposes only scalar parameters for tuning the interfacial width, as discussed in~\cite{moelans2008introduction}. $f_\text{loc}$ is therefore the primary input needed to formulate a new phase-field material model.

In this work, we solve the resulting PDEs using FEM and implicit time integration with the open source Multiphysics Object-Oriented Simulation Environment (MOOSE) \cite{Gaston_2015, tonks_object-oriented_2012}. MOOSE is a finite-element framework primarily developed at Idaho National Laboratory (INL)
which includes several physics modules that assist users in developing phase-field, thermal transport, solid mechanics, and chemistry models.
All the techniques and applications discussed in this work are currently available in the MOOSE \texttt{phase\_field} module.
MOOSE solves systems of PDEs in a tightly-coupled manner by forming a single residual vector comprising all the unknowns.
To form the residual vector, the equations have to be transformed into their \emph{weak form} via multiplication by a suitably defined test function and integration over $\Omega$.
We subsequently utilize the Gauss divergence theorem to reduce the order of the spatial derivatives in the resulting residual equations
(see~\cite{tonks_object-oriented_2012} for more detail on the development of the weak form).

Due to the use of implicit time integration, the system of phase-field equations requires a nonlinear solve at each time step. To solve this system of nonlinear equations, MOOSE typically employs the preconditioned Jacobian-free Newton Krylov~\cite{Brown_1986,knoll:2004} method (PJFNK), via interfaces provided by the libMesh~\cite{libMeshPaper} and PETSc~\cite{petsc-user-ref} libraries. To improve the convergence properties of the nonlinear solve, the preconditioning matrix should be as close as
possible to the actual Jacobian of the nonlinear system of equations.
Computing the Jacobian matrix entries requires derivatives of the residual vector entries with respect to each of the non-linear variables in the system.
This includes, for instance, cross-derivatives of the free energy density functional with respect to all phase-field variables used in the model.

The Cahn--Hilliard Eq.~\eqref{eq:ch_strong} involves a fourth-order spatial derivative on the concentration variable. There are two standard formulations which are commonly used to solve~\eqref{eq:ch_strong}, see, for instance, the discussion in~\cite{zhang_quantitative_2013}. The first is to directly solve the time-dependent fourth-order PDE (using a $C^1$-continuous finite element discretization) and the second is to split the equation into two second-order PDEs which can be solved with more traditional $C^0$ finite elements.
The split solve of the Cahn--Hilliard equation depends on $\partial f_\text{loc}/\partial c_i$, while the the direct solve involves
\begin{eqnarray}
   \nabla \frac{\partial f_\text{loc}}{\partial c_i} = \sum_j \frac{\partial^2 f_\text{loc}}{\partial c_i \partial c_j} \nabla c_j + \sum_k \frac{\partial^2 f_\text{loc}}{\partial c_i \partial \eta_k} \nabla \eta_k.
\end{eqnarray}
If there are $N$ phase-field variables, computing the Jacobian requires $N$ additional second derivatives for the Allen--Cahn equation/split solve of the Cahn--Hilliard equation, and $N^2$ third derivatives for the direct solve of the Cahn--Hilliard equation. The MOOSE \texttt{phase\_field} module
contains FEM discretizations for both the split and non-split Cahn--Hilliard formulations, since there are advantages and disadvantages
to using both~\cite{zhang_quantitative_2013}.

\section{Free energy based approach}
The development of numerical simulation tools for solving phase-field mathematical models can be simplified by employing an approach based on the free energy functional.
In such an approach, the Cahn--Hilliard equation (both the non-split and split forms) and the Allen--Cahn equation obtain all the required free energy density derivatives
from free energy objects (in the sense of object-oriented programming) which are specialized to the application in question. When a new phase-field model is developed, the existing implementations of the residual equations are re-used---only the code required to define the specialized free energy object itself must be written.

In the MOOSE \texttt{phase\_field} module, Cahn--Hilliard and Allen--Cahn residuals and Jacobians have been implemented using small program units called ``Kernels''. A Kernel is a \cpp{} class representing a term in the weak form of a PDE. The computation of the free energy density and its derivatives is performed in MOOSE by ``Material'' objects. A Material computes values, which may depend on non-linear variables, at specified points in the simulation domain. All simulation-specific physics are implemented through Material objects which calculate the free energy density. The free energy density derivatives are then used in a generic way by the Kernel objects.

\section{Automatic differentiation of free energies}
While the free energy based approach to phase-field model implementation simplifies the overall process, manually programming the free energy derivatives can be both burdensome and a potential source of error. Therefore, automatic differentiation is used to simplify the process even further, requiring the user to implement only the free energy density expression itself. All required derivatives are computed analytically in a fully-automated way.

\begin{figure}[tbp]
\centering
\begin{tcolorbox}[width=.7\linewidth]
  \centering
  \epsfig{width=\linewidth,file=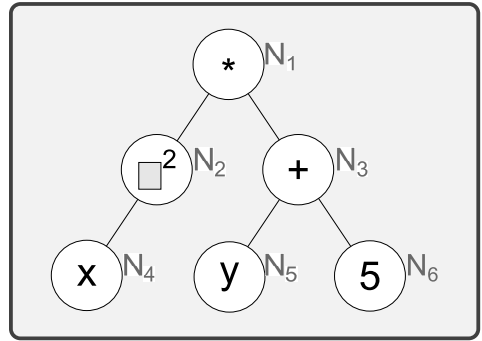}
\end{tcolorbox}
  \caption{\label{fig:tree} Schematic example of the tree representing the mathematical expression $x^2(y+5)$. The nodes $N_1$ and $N_3$ represent the multiplication and sum operators, respectively, with two arguments each, the internal node $N_2$ represents a square function with one argument, and the leaf nodes $N_4$--$N_6$ represent the variables $x$, $y$, and the constant $5$, respectively.}
\end{figure}

To facilitate automatic differentiation, and to allow user-defined functions to be supplied via input files, MOOSE employs the Function Parser library~\cite{fparser-web-page} that is included as a third-party plugin in the underlying libMesh finite element library. The Function Parser library accepts a mathematical function definition given as a plain text string.
The expression string is lexically parsed into an intermediate tree representation and then transformed into stack machine bytecode. This bytecode can then be executed by the Function Parser bytecode interpreter module as often as necessary without further transformation.
This intermediate tree representation of the Function Parser expressions, illustrated in Fig.~\ref{fig:tree}, readily lends itself to algorithmic transformations such as automatic differentiation.

The automatic differentiation system, which is a major new contribution in this work, operates on the tree representation of the free energy function. In this tree structure, leaf nodes can correspond to constants or variables, and internal nodes correspond to mathematical operators and functions, the arguments of which are contained in child nodes or subtrees. The derivatives of the leaf nodes are $0$ for all nodes that do not represent the variable the derivative is taken with respect to, and $1$ for all the nodes that do. The derivatives of the internal nodes are constructed recursively according to a set of elementary derivative rules.

Construction of the derivative starts at the root node of the expression tree. For the example expression tree in Fig.~\ref{fig:tree}, which represents the expression \mbox{$x^2(y+5)$}, the root node holds the multiplication $N_1=N_2\ast N_3$. To obtain the derivative with respect to $x$, we need to calculate the derivative of the root node, $d_xN_1$. We set $d_xN_1 = d_xN_2\ast N_3 + N_2\ast d_xN_3$ according to the product rule.
This expression contains derivatives of the nodes $N_2$ and $N_3$.  These derivatives are recursively constructed until leaf nodes with a zero derivative value are reached.  This happens in all cases except $d_xN_4$, which evaluates to one. The full derivative expression that is constructed in this manner is $(2x\ast 1)\ast(y+5) + x^2(0+0)$.

Our method is a member of the class of source transformation or symbolic differentiation algorithms~\cite{TOLSMA1998475,Kedem1980}. In contrast to the commonly used forward and backward accumulation automatic differentiation algorithms, the derivative evaluation is not tied to the evaluation of the undifferentiated function. Each derivative is algebraically optimized, compiled, and evaluated exactly when needed. In a finite element code, the weak form residual evaluation (which, in a split form Cahn--Hilliard problem, contains the first derivative of the free energy only) is the most common operation by far, occurring at every linear iteration. The Jacobian, containing the second derivatives, is only evaluated once per non-linear iteration. The undifferentiated form is not evaluated for the purpose of solving the phase-field equations, but only when needed for output or postprocessing stages, which is generally once per time step. MOOSE furthermore detects which derivatives are needed by the residual and Jacobian evaluations and skips evaluation of unused derivatives, which reduces the computational burden when constructing approximate Jacobian matrices for preconditioning purposes.

The Function Parser library provides a comprehensive algebraic optimizer that groups, reorders, and transforms the function expression into an equivalent but faster-to-evaluate form. The algebraic simplifications are essential for removing the trivial leaf node derivatives which may lead to evaluation errors, such as division by zero, and can be avoided by simple term cancellations. In the above example, the simplifications reduce the derivative expression to $2x(y+5)$.

To further improve the performance of the parsed and runtime interpreted functions, we have also developed a just-in-time (JIT) compilation module.
At runtime, the generated bytecode sequences are automatically transformed into small C source code files. A compiler is dispatched to compile each function file into a dynamically linkable library, which is then loaded ``on the fly'' with a \texttt{dlopen}~\cite{dlopen-page} POSIX system call. This occurs once during simulation initialization. If the JIT compilation fails, the function evaluation falls back on the bytecode interpreter, otherwise the generated machine code is called.
The average time overhead of the additional compilation step is on the order of 0.1 s per function expression or less, depending on the system the simulation is executed on. This is further mitigated by a caching system. A unique SHA1 hash~\cite{sha1} is computed from the function bytecode, and the compiled functions are stored permanently using the hash as a filename. Recompilation will only occur if the bytecode, and thus the function expression, changes. Trivial function changes, namely the modification of constants, will in most cases not trigger a recompilation.

\begin{figure}
  \centering
  \epsfig{width=0.9\linewidth,file=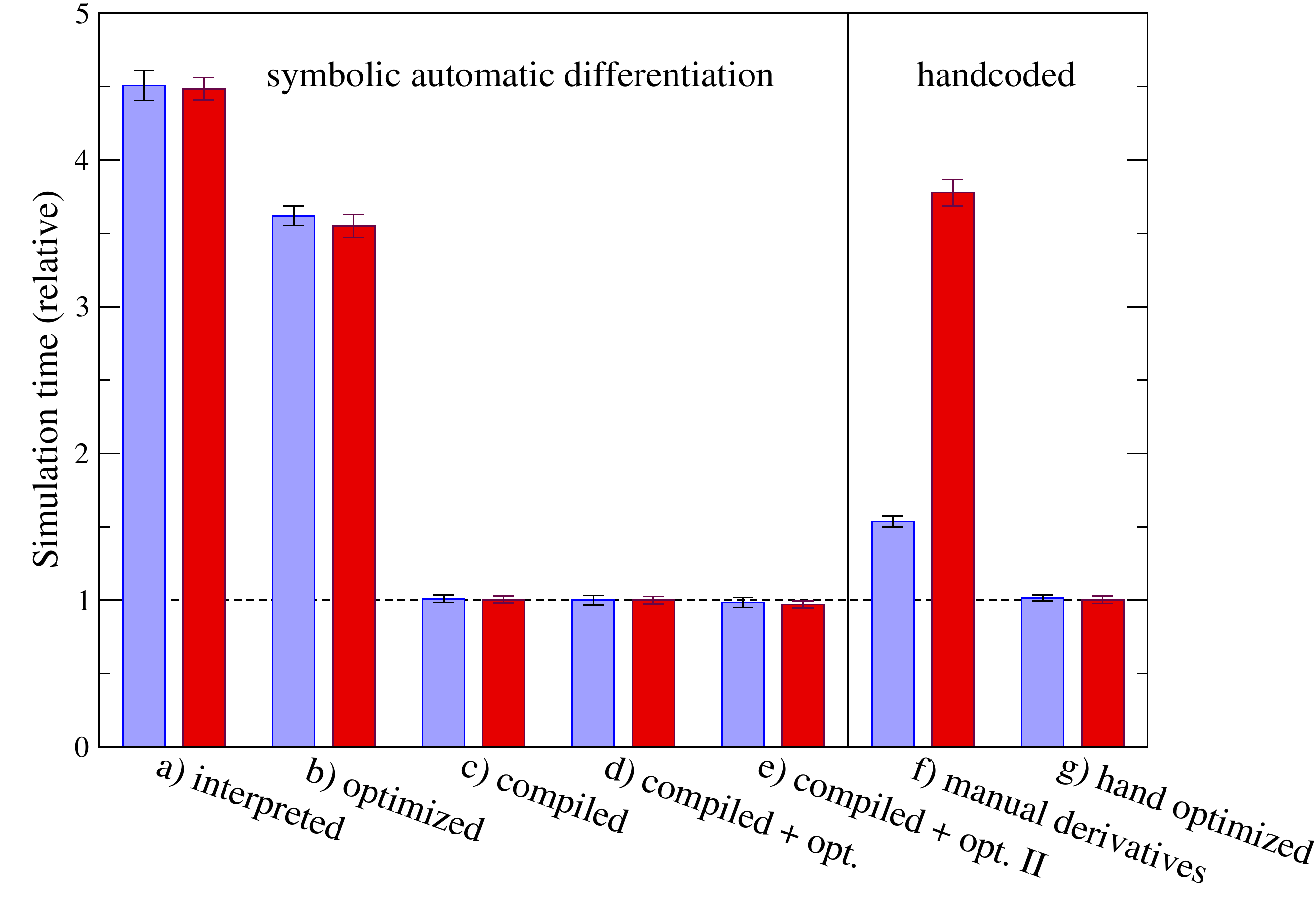}
  \caption{\label{fig:jitperf} Relative computation times for an iron chromium phase-field simulation. Hand coded free energy derivatives are compared to derivaties computed via the automatic symbolic differentiation system. Results a) and b) use the standard Function Parser bytecode interpreter, results c), d) and e) use the just-in-time compilation (JIT) module developed in this work. Results b), d), and e) also use the algebraic optimization provided by FParser. Blue (light gray) bars are obtained using the clang compiler, red (dark gray) results are obtained with GCC.}
\end{figure}

In Fig.~\ref{fig:jitperf}, the run times for an iron chromium phase-field simulation (described in detail below) are compared for different implementations of the free energy. Results (a) through (e) are obtained using our symbolic differentiation system, while results (f) and (g) are obtained using hand coded derivatives. The outputs generated by those runs have been confirmed to be identical within the limits of the convergence tolerance. The blue (light gray) bars were obtained using the LLVM-based Clang compiler~\cite{Lattner2004,Clang} (version 3.9.0), and the red (dark gray) results were obtained using the GNU Compiler Collection~\cite{GCC} (GCC) \cpp{} compiler (version 6.2.0). Results (a) and (b) employ the standard Function Parser bytecode compiler. Result (b) shows the effect of applying the built-in algebraic optimization to the function byte code. Results (c), (d) and (e) use the JIT compilation module developed for this work. JIT compilation turns out to be the crucial step needed to ensure performance on par with hand-coded derivatives as shown in result (g).

The JIT module has been incorporated into the Function Parser library fork in libMesh. Through this automatic symbolic differentiation system, we achieve a significant reduction in developer time and remove a source of developer error that is difficult to track down and debug.
We note that a naive implementation of the free energy obtained via manual differentiation can exhibit significantly worse performance. One factor is the use of the \cpp{} library function \texttt{std::pow}, which for integer powers is significantly slower than repeated multiplication operators. The Function Parser library performs this optimization automatically.

\begin{figure}[tbp]
  \centering
  \epsfig{width=0.9\linewidth,file=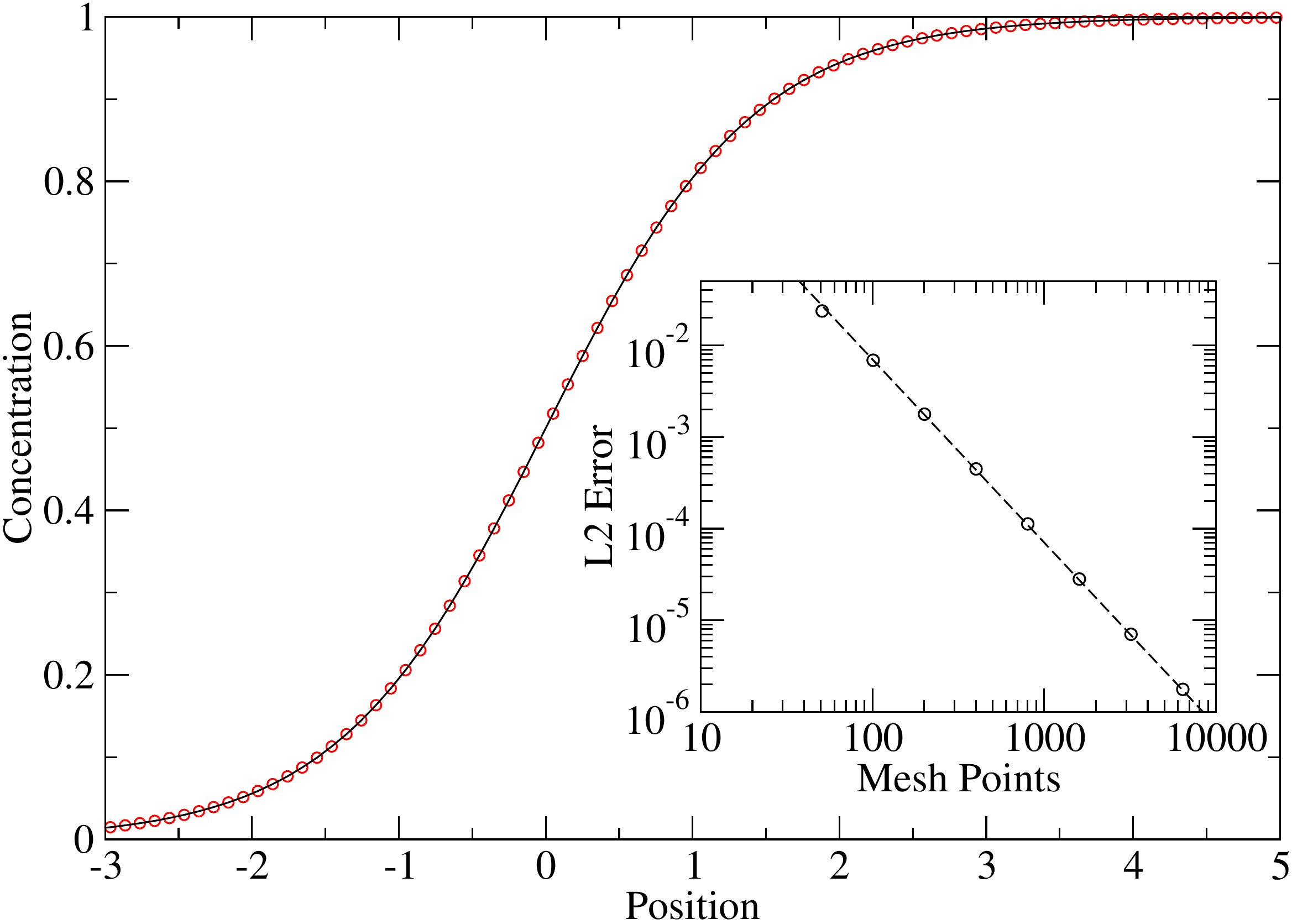}
  \caption{\label{fig:analytic} Comparison of the numerical solution (red circles) of the Cahn--Hilliard equation in the split form on linear Lagrange elements and the analytical solution (solid line) for an equilibrium interface in a one dimensional domain. The inset shows that the $L^2$ norm of the error in the numerical solution
converges at $O(h^2)$, where $h$ is a measure of the grid size.}
\end{figure}

To demonstrate the correctness of the implementation, we compare the numerical solution of a Cahn--Hilliard equation for a simple double well free energy given by $f=c^2(1-c)^2$ (interface parameter equal to 1) to the analytical solution, $c=\frac12\left(1+\tanh\left(\frac x{\sqrt 2}\right)\right)$ for the equilibrium interface in a one dimensional domain. Fig.~\ref{fig:analytic} shows good agreement of the interface profiles obtained numerically (red circles) and the analytical solution (solid line). The inset shows the expected quadratic convergence with respect to the mesh resolution for linear Lagrange elements. The split formulation of the Cahn--Hilliard equation was used along with automatic differentiation of the free energy. To the authors' best knowledge this integrated approach of utilizing run time symbolic differentiation and JIT compiling of free energies in a phase-field framework is unique.

\subsection{Smoothly-extrapolated logarithm}
Free energies that contain a term for the configurational entropy derived from ideal or regular~\cite{Hildebrand} solution models will contain terms of the form
\begin{equation*}
c\ln c + (1-c)\ln(1-c),
\end{equation*}
where $c$ is a conserved order parameter. As the natural logarithm is only defined for strictly positive numbers, this expression restricts the domain of the free energy to numbers on the open interval $(0,1)$. This poses numerical challenges for systems with equilibrium concentrations near 0 or 1.

To improve the convergence behavior, we have developed a smoothly-extrap\-olated logarithm surrogate function. For input arguments $c<\varepsilon$, we compute a Taylor-series expansion of the logarithm function centered around $c=\varepsilon$ instead. For $c > \varepsilon$, we evaluate the standard logarithm function. This surrogate logarithm function also extends to negative arguments, and is twice differentiable everywhere. In the resulting free energy expressions, the extension to negative arguments manifests as a free energy penalty which drives the solution back to physically-allowable concentrations without incurring numerical errors.
In previous work~\cite{Jokisaari2015334} piecewise constructions using Taylor expansions of the full free energy expressions outside its domain have been suggested. This requires knowledge of the domain of the free energy to set the interval boundaries over which the Taylor expansion is active. Our approach retains a single free energy expression with the Taylor expansion only occurring at the level of the logarithm functions, which are the underlying cause of the limitation of the domain of the free energy. The domain of the log function is well defined and thus our model requires fewer user inputs.

Care has to be taken to choose $\varepsilon$ small enough to not adversely impact the thermodynamic properties of the simulated system. In particular, large values of $\varepsilon$ can change the phase diagram by moving the common tangent points to larger concentrations.

\begin{figure}
  \centering
  \epsfig{width=0.7\linewidth,file=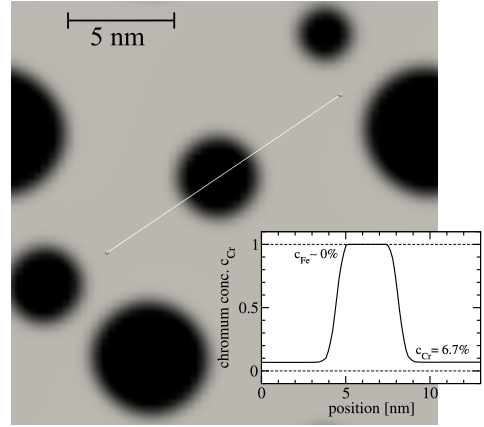}
  \caption{\label{fig:fecr} Snapshot of a phase-field simulation of spinodal decomposition and formation of chromium rich precipitates in an iron chromium alloy obtained using a runtime parsed and automatically differentiated free energy expression. The line scan plotted in the inset shows the precipitate and matrix concentrations.}
\end{figure}
As an example, we used the techniques presented above to implement a phase-field model based on the published~\cite{schwen_analytic_2013} free energy surface of an iron chromium binary alloy. We have encoded the full free energy expression from the publication into a MOOSE input file as a parsed function Material with automatic differentiation and Taylor expansion substitution for small logarithm arguments. A running phase-field model was obtained with little additional effort. Fig.~\ref{fig:fecr} shows a simulation snapshot obtained using this free energy. The system is in the particle coarsening stage, having previously undergone spinodal decomposition. A line scan performed on the center precipitate is shown in the inset. In agreement with the published free energy surface and resulting phase diagram, we observe practically no solubility of iron in the chromium precipitates, while the chromium solubility in the iron matrix is approximately 6.7\% at the simulation temperature of 500\,K.

In addition to the free energy, the user has to provide a mobility model, which for this example we assumed to be concentration independent.
We used experimental data on chromium tracer diffusivity in iron as a guide to set the actual value of the mobility $M =10^{-3}~\text{nm\,(eV\,s)}^{-1}$. For the chosen simulation length scale of 20 nm $\times$ 20nm and mesh of $3600$ quadrilateral elements, an appropriate interfacial energy parameter $\kappa = 0.3$ eV\,nm$^{-1}$ was chosen. The initial choice of $\kappa$ is somewhat arbitrary, as its order of magnitude rarely changes at a given length scale, and common excess free energies are of similar magnitude for many alloy systems.
Further refinement of the $\kappa$ value to obtain a specific surface energy may require a few simulation trial runs, which on one-dimensional test systems take only seconds.
The example simulation was run with an adaptive time stepper to a simulation time of about 24 h.
All input files required to rerun the simulations in this paper are available from the MOOSE GitHub repository\footnote{\url{https://github.com/idaholab/moose/tree/devel/modules/combined/examples/publications/rapid_dev}}.

\section{Multiphase phase-field models}

The tools presented above allow the rapid implementation of single phase material systems, however, many material systems of interest exhibit complex phase diagrams with multiple phases potentially coexisting in a simulation volume. These phases have separate free energies and can potentially have different mechanical and thermal properties.
To model systems with multiple coexisting phases, the construction of a global free energy functional spanning the entire phase space of the system is necessary.
We now show that the material-based modular free energy system presented here lends itself to the convenient construction of such multiphase free energies.
The free energy of each individual phase can be provided using a Material object that encapsulates all required free energy derivatives and leverages the capabilities of the symbolic differentiation module.
The global free energy is then constructed as a combination of the phase free energies using non-conserved order parameters to indicate the phase distribution. Just as the free energy values are combined, the derivatives for the phase free energies are used to construct the derivatives of the global free energy.
The MOOSE phase-field module has a selection of pre-made Material objects for global free energy construction which users can provide single phase free energies for.

One common approach is to use a linear combination of the free energy densities $f^j_{loc}$ of each phase in the system based on the WBM model~\cite{wheeler1992phase}, e.g.
\begin{equation}\label{eq:wbm}
  f_\text{loc}=\sum_j h(\eta_j)f_\text{loc}^j(c_1,c_2,\ldots) + Wg(\eta_1,\eta_2,\ldots).
\end{equation}
The switching function $h(\eta_j)$ varies smoothly from $0$ to $1$ as $\eta_j$ goes from $0$ to $1$. The barrier function $g$ (multiplied by the barrier height $W$) penalizes phase mixtures over pure phase regions. The total weight of all phase free energy contributions at each point in the simulation volume is exactly unity, which can be expressed as the constraint
\begin{equation}
  \label{eq:kdef}
  k(\eta_1,\eta_2,\ldots) \equiv \sum_j h(\eta_j) - 1 = 0.
\end{equation}

Two phase systems can be modeled using a single order parameter $\eta_1$ with the explicit constraint $\eta_2=1-\eta_1$. The symmetric switching function $h(\eta)=1-h(1-\eta)$ then satisfies the constraint~\eqref{eq:kdef}. For $n$-phase systems with $n>2$, it is advantageous to use $n$ order parameters. In this case, the constraint $k$ is not automatically satisfied and needs to be enforced by other means. In the MOOSE phase-field module we provide two methods of enforcing the switching function constraint: a ``hard'' constraint utilizing the Lagrange multiplier technique, and a ``soft'' constraint implemented via a free energy penalty term.

The soft constraint is implemented by adding the contribution,
\begin{equation}
  f_p = \chi \left[1 - \sum_j h(\eta_j)\right]^2,
\end{equation}
to the free energy, where $\chi$ is a user-tunable penalty coefficient.
In contrast, the hard constraint is imposed by introducing a Lagrange
multiplier $\lambda$ as a field variable.  The variational statement of the
problem is then: find $(\eta_1,\eta_2,\ldots, \lambda)$ such that the boundary
conditions are satisfied, and
\begin{align}
  \label{eq:L1res}
  a_j(\eta_1,\eta_2,\ldots, v) + \int_\Omega\lambda\frac{\partial k}{\partial\eta_j} v\,\text{d}x &= 0
  \\
  \label{eq:L2res}
  \int_\Omega q\frac{\partial(\lambda k)}{\partial\lambda}\,\text{d}x &= 0
\end{align}
hold for all admissible test functions $(v,q)$, where
$a_j(\eta_1,\eta_2,\ldots, v)$ is the weak form
(Allen--Cahn) residual for the $j$th non-conserved order parameter.
We note that these equations alter the character of the Jacobian matrix of the non-linear problem by introducing a zero block on the diagonal. This can complicate the task of preconditioning and iteratively solving the system substantially. By replacing the constraint $k$ with the modified constraint
\begin{equation}
  \bar k(\eta_1,\eta_2,\ldots, \lambda) \equiv k(\eta_1,\eta_2,\ldots) - \frac\varepsilon2\lambda,
\end{equation}
the $\frac\varepsilon2\lambda$ term introduces an $\mathcal{O}(\varepsilon)$ $\lambda$-dependence in the constraint.
This results in a non-zero on-diagonal Jacobian contribution of $-\varepsilon$ for Eq.~\eqref{eq:L2res}, avoiding
``zero pivot'' errors arising from PETSc preconditioners (such as \texttt{-pc\_type lu}, which implements only partial pivoting).
The value of $\varepsilon$ should be chosen slightly larger than the linear solver tolerance, and results in a trade off between accuracy and solver performance.
This approach does result in a violation of the constraint of $\mathcal{O}(\varepsilon)$, however it was found that
this discrepancy did not adversely affect the overall quality of the solution, and improved
the convergence characteristics of the solver.

%
%
\subsection{KKS models}
\begin{figure}[tbhp]
  \centering
  \epsfig{width=0.9\linewidth,file=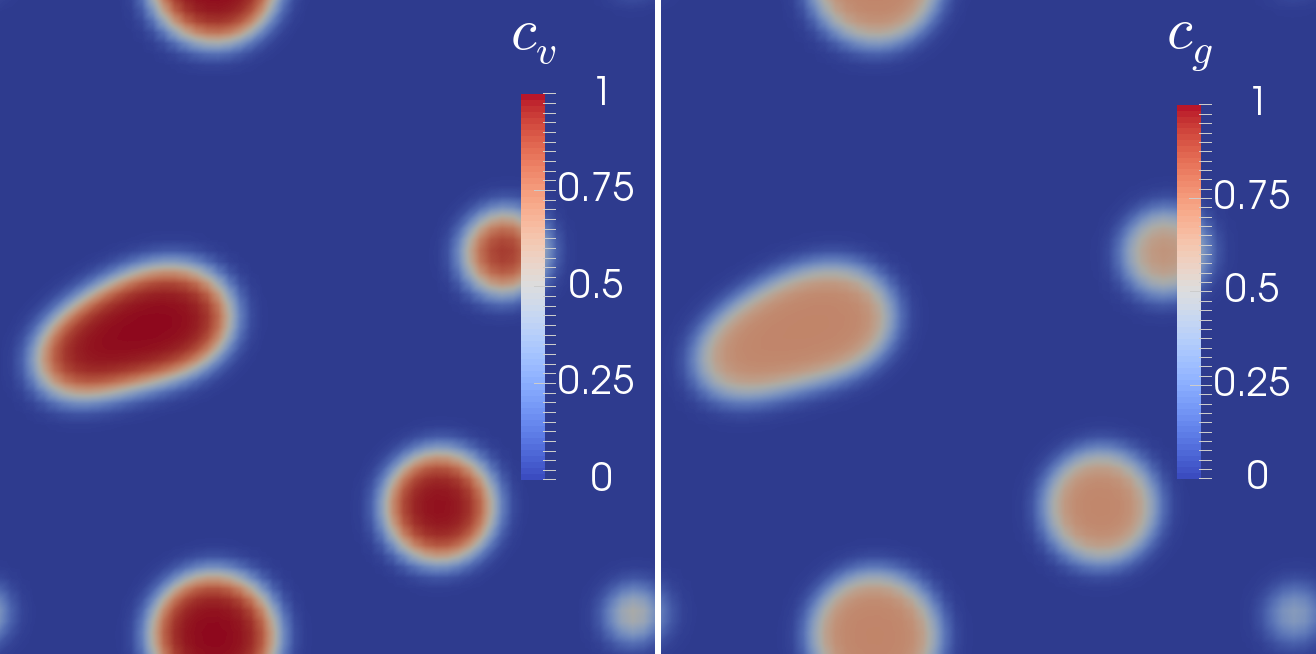}
	\caption{\label{fig:KKS_UO2_Xe}Example of bubble formation modeled with the KKS UO$_2$ fission gas model, where the vacancy concentration is shown on the left and the Xe concentration is shown on the right. The domain is 10 nm on each side with periodic boundary conditions applied.}
\end{figure}

An additional multiphase model implemented in the \texttt{phase\_field} module is the Kim-Kim-Suzuki (KKS) model~\cite{kim_phase-field_1999}. KKS addresses the issue of systems with large phase free energy differences in the interfacial regions.
One relevant example is the xenon gas bubble problem shown in Fig.~\ref{fig:KKS_UO2_Xe}, computed using the phase free energies from Li et al.~\cite{li_phase-field_2013}. Here the gas solubility in the solid UO$_2$ matrix is very low, with large free energy penalties for large gas concentrations. In the bubble phase, the equilibrium gas concentration is near unity. In the bubble/matrix interfacial region, both the order parameter and the concentration change from the bubble equilibrium values to the matrix equilibrium values over a finite distance due to the soft interface approximation. In that interfacial region, the free energy of both phases is computed for the intermediate concentration range, which results in large free energy densities from the solid phase contribution. This increases the interfacial free energy of the bubbles to a nonphysical value.

The KKS model solves this problem by introducing the concept of phase concentrations, which are effectively the fractions of the total concentration held in a given phase. In this model, the gas concentration is largely shifted to the gas phase to avoid the free energy penalty. In the KKS model, the interfacial free energy is decoupled from the diffuse interface width, allowing the use of wider interfaces and therefore coarser spatial discretizations  requiring fewer computational resources. However, solving for these new variables requires additional differential equations. In the KKS model, these are given by requiring mass conservation and equality of the component chemical potentials across all phases:
\begin{align}
  F &= \left(1-h(\eta)\right) F_a(c_a) + h(\eta)F_b(c_b) + Wg(\eta),
  \\
  c &= \left(1-h(\eta)\right)c_a + h(\eta)c_b,
  \label{eq:KKSConstraint}
  \\
  \frac{\partial f_a}{\partial c_a} &= \frac{\partial f_b}{\partial c_b}.
\end{align}

The MOOSE phase-field module currently implements a two-phase version of the KKS model that uses Kernels for the phase-field equations as well as the KKS constraint equations. The free energy is supplied to those Kernels using the derivative Material system outlined above in the same manner as for the other multiphase approaches discussed above.

In addition to the two-phase KKS model, a three-phase model based on~\cite{Ohno2010} has been implemented in the MOOSE phase-field module. One limitation of many multiphase-field models is that a binary interface between two phases is unstable with respect to the spurious formation of additional phases in the diffuse interfacial region~\cite{Folch2005, Toth2015}.
The formation of the spurious third phase occurs when the free energy is not convex with respect to that third phase. Its formation distorts the interfacial energy of the binary interface, and can lead to nucleation of the third phase in unphysical locations. This can be mitigated by requiring the switching functions to have the additional property of having zero slope and positive curvature perpendicular to the transformation path between two phases. Functions with that property are called \emph{tilting functions}. In the three-phase model implemented in MOOSE, the tilting functions developed in~\cite{Folch2005} and applied to the KKS approach in~\cite{Ohno2010} are used to prevent spurious third-phase formation at a two-phase interface.

In this model, the three phases are represented by order parameters $\eta_1$, $\eta_2$, and $\eta_3$.  They are constrained such that
\begin{equation}
  \label{eq:constraint}
  \eta_1 + \eta_2 + \eta_3 = 1
\end{equation}
The free energy of the system is given by
\begin{equation}
F = \int_\Omega \left( f_\text{loc} + f_\text{gr}  \right) \text{d}V
\label{eq:energy}
\end{equation}
The local energy $f_\text{loc}$ is given by
\begin{equation}
f_\text{loc} = \sum_{i=1}^3 \left[ h_i f_i(c_i) + W \eta_i^2 (1-\eta_i)^2 \right]
\label{eq:floc}
\end{equation}
where $f_i$ is the free energy density of each phase, $c_i$ is the phase concentration, $W$ is the potential barrier height, and the $h_i$ are the tilting functions~\cite{Folch2005, Ohno2010}
\begin{align}
  \nonumber
  h_i &\equiv h_i(\eta_i, \eta_j, \eta_k)
  \\
  \label{eq:tilting}
                  &=  \frac{\eta_i^2}{4} \left(15(1 - \eta_i) [1 + \eta_i - (\eta_k - \eta_j)^2] + \eta_i(9\eta_i^2 - 5)\right)
\end{align}
for $i=1,2,3$, and where
\begin{align}
 j &\equiv 1 + \text{mod}(i,3) \\
 k &\equiv 1 + \text{mod}(i+1,3)
\end{align}
i.e.\ $(i, j, k)$ form a cyclic permutation. The gradient energy $f_\text{gr}$ is given by
\begin{equation}
f_\text{gr} = \sum_i \frac{\kappa}{2} \left| \nabla \eta_i \right|^2
\label{eq:fgrad}
\end{equation}
The concentration $c$ is defined as a function of the phase concentrations as
\begin{equation}
c = \sum_{j=1}^3 h_j c_j
\label{eq:c}
\end{equation}
Because the tilting function $h_i$ reduces to the commonly used two-phase interpolation function $h_i = \eta_i^3(10-15\eta_i + 6\eta_i^2)$ along the two-phase interfaces~\cite{Folch2005}, this constraint on the concentrations reduces to~\eqref{eq:KKSConstraint}, the constraint in the two-phase KKS model, for $i-j$ interfaces. The phase concentrations are constrained such that the chemical potentials of each phase are equal:
\begin{equation}
\mu = \frac{\partial f_1}{\partial c_1} = \frac{\partial f_2}{\partial c_2} = \frac{\partial f_3}{\partial c_3}
\label{eq:chempot}
\end{equation}

To enforce the constraint of~\eqref{eq:constraint}, a term $\lambda \left(1-\sum_i \eta_i \right)$ is added to the free energy functional, where $\lambda$ is a Lagrange multiplier. This results in the Lagrangian $F_L$:
\begin{equation}
F_L = \int_\Omega \left[ f_\text{loc} + f_\text{gr} + \lambda \left(1-\sum_i \eta_i \right) \right] \text{d}V
\label{eq:energy_lagrangian}
\end{equation}

The time evolution of the order parameters is given by the Allen--Cahn equation
\begin{equation}
\frac{\partial \eta_i}{\partial t} = -L \frac{\delta F_L}{\delta \eta_i}
\label{eq:AllenCahn}
\end{equation}
where using the variational derivative of the Lagrangian enforces the constraint~\eqref{eq:constraint}. Using the fact that $\frac{\partial}{\partial t} \sum_i \eta_i = 0$ and assuming the mobilities for each phase are equal at each position (though they may be dependent on the local values of the order parameters), the Lagrange multiplier can be eliminated and the Lagrangian can be written in terms of the non-constrained variational derivatives as
\begin{equation}
\frac{\delta F_L}{\delta \eta_i} \bigg|_{\sum \eta_i = 1} = \frac{\delta F}{\delta \eta_i} - \frac{1}{3} \sum_j \frac{\delta F}{\delta \eta_j}
\label{eq:variational}
\end{equation}
Substituting for $F$,
\begin{align}
  \nonumber
  \frac{\partial \eta_i}{\partial t} = &-L \left[ \frac{2}{3}\left( \frac{\partial f_\text{loc} }{\partial \eta_i} - \nabla \cdot \frac{\partial f_\text{gr} }{\partial \nabla \eta_i} \right) \right.
  \\
  \nonumber
  &\qquad -\frac{1}{3}\left( \frac{\partial f_\text{loc} }{\partial \eta_j} - \nabla \cdot \frac{\partial f_\text{gr} }{\partial \nabla \eta_j} \right)
  \\
  \label{eq:AllenCahn2}
  &\qquad -\left. \frac{1}{3}\left( \frac{\partial f_\text{loc} }{\partial \eta_k} - \nabla \cdot \frac{\partial f_\text{gr} }{\partial \nabla \eta_k} \right) \right]
\end{align}
where
\begin{align}
  \nonumber
  \frac{\partial f_\text{loc}}{\partial \eta_i} &= \frac{\partial h_i}{\partial \eta_i} [f_i(c_i) - \mu c_i] \
  \\
  \nonumber
  &+ \frac{\partial h_j}{\partial \eta_i} [f_j(c_j) - \mu c_j]
  \\
  \nonumber
  &+ \frac{\partial h_k}{\partial \eta_i} [f_k(c_k) - \mu c_k]
  \\
  \label{eq:flocloc}
  &+ 4 W \eta_i \left(\eta_i-1\right) \left(\eta_i-\frac{1}{2}\right)
\end{align}
and
\begin{equation}
\nabla \cdot \frac{\partial f_\text{gr} }{\partial \nabla \eta_i}\label{eq:fgrad2} = \kappa \nabla^2 \eta_i
\end{equation}

The time evolution of the concentration field is given by the Cahn--Hilliard equation:
\begin{equation}
\frac{\partial c}{\partial t} = \nabla \cdot \left[ M_c \nabla \frac{\delta F}{\delta c} \right]
\label{eq:CahnHilliard}
\end{equation}
where $M_c$ is the Cahn--Hilliard mobility. Using the chain rule, it can be shown following Kim et al.~\cite{GyoonKim2004135} that this is equivalent to
\begin{equation}
\frac{\partial c}{\partial t} = \nabla \cdot \left[ D \sum_i h_i \nabla c_i \right]
\label{eq:diffusion}
\end{equation}
where $D$ is the solute diffusion coefficient. This demonstrates that the evolution equation can be solved in terms of the phase concentrations.

\begin{figure}[hbt]
  \centering
  \epsfig{width=0.7\linewidth,file=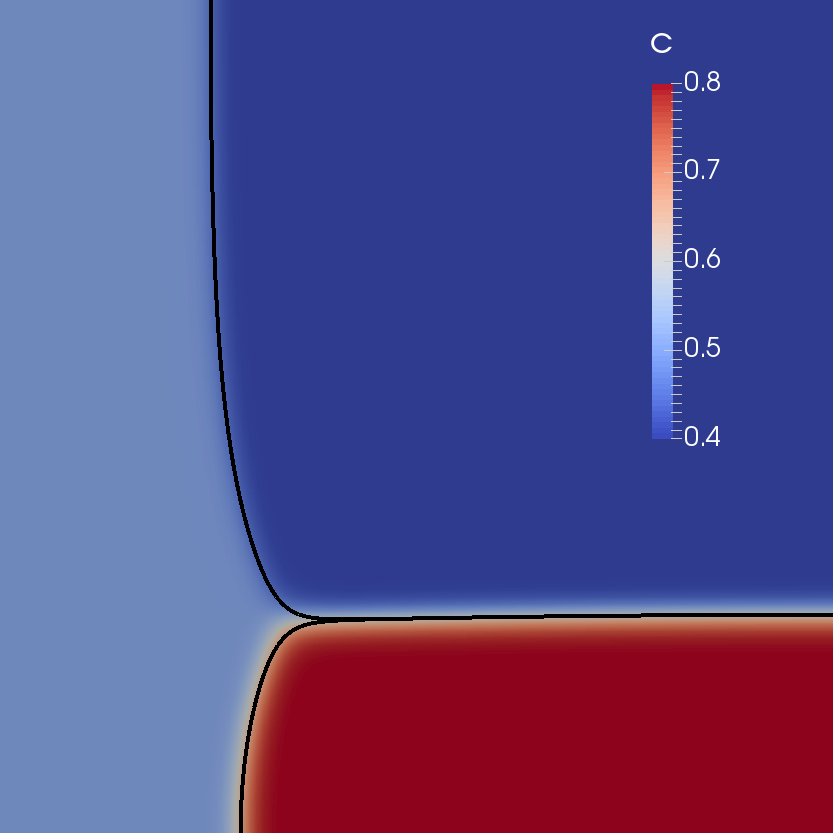}
  \caption{\label{fig:eutectic} Tri-junction formed by a three-phase system at the eutectic composition, with equilibrium compositions of each phase $c_1 = 0.4$, $c_2 = 0.5$, and $c_3 = 0.8$, simulated using the MOOSE three-phase KKS model.  The simulation cell is 40\,nm $\times$ 40\,nm. The boundaries between phases are indicated with contour lines at $\eta_1 = 0.5$ and $\eta_3 = 0.5$.}
\end{figure}

The three-phase KKS model was used to simulate a tri-junction in a simple model system, as shown in Fig.~\ref{fig:eutectic}. The three phases have parabolic free energies, supplied using the derivative Material system, with minima at $c_1 = 0.4$, $c_2 = 0.5$, and $c_3 = 0.8$. The free energy of phase 2 was made temperature-dependent, and a fixed temperature gradient was imposed to stabilize the tri-junction. As seen in Fig.~\ref{fig:eutectic}, at equilibrium a stable tri-junction is formed. The boundaries between phases are indicated with contour lines at $\eta_1 = 0.5$ and $\eta_3 = 0.5$. No third-phase formation is observed at any two-phase interface. It was also verified  separately that the order parameter and composition profiles through a 1D $\eta_1 -\eta_2$ interface match the analytical solutions, $\eta_1 = \frac{1}{2}\left[ 1 - \tanh{\frac{x - x_0} {\sqrt{2} \delta}} \right]$, $\eta_2 = 1- \eta_1$, $\eta_3 = 0$, and $c = h_1(\eta_1, \eta_2, 0) c_1 + h_2(\eta_1, \eta_2, 0) c_2$ (where $x_0$ is the midpoint of the interface and $\delta = \sqrt{\frac{\kappa}{W}}$), and that the interfacial energy matched the analytical solution, $\gamma = \frac{\sqrt{2 \kappa W}}{3}$. Thus, the three-phase KKS model implemented in the MOOSE phase-field model allows simulation of three-phase systems with arbitrary free energies (specified as before with the derivative Material system), prevents third-phase formation at a two-phase interface, decouples interfacial energy from local (chemical) energies, and allows the interfacial energy and thickness to be set independently for optimal efficiency.

\section{Mechanics coupling}
Heterogeneous material properties in multiphase simulations, such as lattice mismatches and variations in the elasticity tensor, will introduce an interplay between chemical and mechanical driving forces. Thus, it is critical to couple the phase-field model equations to equations defining the mechanical behavior of the material. Mechanics simulations are available in the MOOSE \texttt{tensor\_mechanics} module, which can be easily coupled to the \texttt{phase\_field} module. The local displacement vector $\vec{u}$ is determined by solving the stress divergence equation
\begin{eqnarray}
  \nabla \cdot \boldsymbol{\sigma}(\boldsymbol{\epsilon} - \boldsymbol{\epsilon^*}) + \vec{b} = 0,
\end{eqnarray}
where $\boldsymbol{\sigma}$ is the stress and $\vec{b}$ is an applied body force. The system is solved given suitable boundary conditions and a constitutive law defining the relationship between stress and the strain, $\boldsymbol{\epsilon}$. A stress free strain (or eigenstrain) $\boldsymbol{\epsilon^*}$ accounts for lattice mismatches, thermal expansion, etc. The elastic energy of the system
\begin{eqnarray}
  E_{el} = \frac{1}{2} \boldsymbol{\sigma}(\boldsymbol{\epsilon} - \boldsymbol{\epsilon^*}) \cdot (\boldsymbol{\epsilon}- \boldsymbol{\epsilon^*})
\end{eqnarray}
is added to the total free energy to account for its impact on the microstructure evolution.

In multiphase models, two approaches exist to model the elastic free energy of the total system:
\begin{enumerate}
\item In the Voight-Taylor scheme~\cite{Ammar2009} each phase has its own mechanical properties, including the elastic constants, constitutive model (and stress), and eigenstrain, which can depend on composition variables. Each phase free energy contains an elastic free energy contribution. The global elastic free energy is computed analogously to the total chemical free energy, i.e.
\begin{eqnarray}
  E_{el} = \sum_j h(\eta_j)E_{el}^j(\boldsymbol{\mathcal{C}}_j, \boldsymbol{\epsilon}^*(\vec{c})), \label{eq:Eel_comb}
\end{eqnarray}
and the total stress is calculated in a similar manner
\begin{eqnarray}
  \boldsymbol{\sigma} = \sum_j h(\eta_j) \boldsymbol{\sigma_j}.
\end{eqnarray}
\item In the Khachaturyan scheme~\cite{Khachaturyanbook, Vaithyanathan2002, Vaithyanathan2004} only global interpolated mechanical properties, which can depend on phase variables, are computed. In this model, only a single stress is computed, and only a single elastic free energy is built, which is added to the total chemical free energy.
\end{enumerate}

In both cases, it is necessary for the mechanical property materials to provide derivatives with respect to their dependent variables. A solute concentration dependent eigenstrain Material, for example, provides the $\frac{\partial\epsilon^*}{\partial c}$ derivative.
The Function Parser system currently only operates with scalar quantities. Free energy contributions requiring vector or tensor terms need to be implemented separately. MOOSE provides a component that computes linear elastic energy contributions to the free energy. To maintain a modular free energy approach, we have created materials that define the elastic contributions to the free energy and its derivatives. In the first approach, this elastic contribution is computed for each phase individually, while in the second approach only one elastic free energy computation is performed in the system.
Additional material classes combine the energy contributions according to Eq.~\eqref{eq:Eel_comb}. A utility Material is then used to add the chemical and mechanical free energies into a total phase free energy. Mechanical material properties with a direct dependence on the phase-field variables will automatically yield coupling terms between the mechanics and phase-field equations.

\begin{figure}
  \centering
  \epsfig{width=0.9\linewidth,file=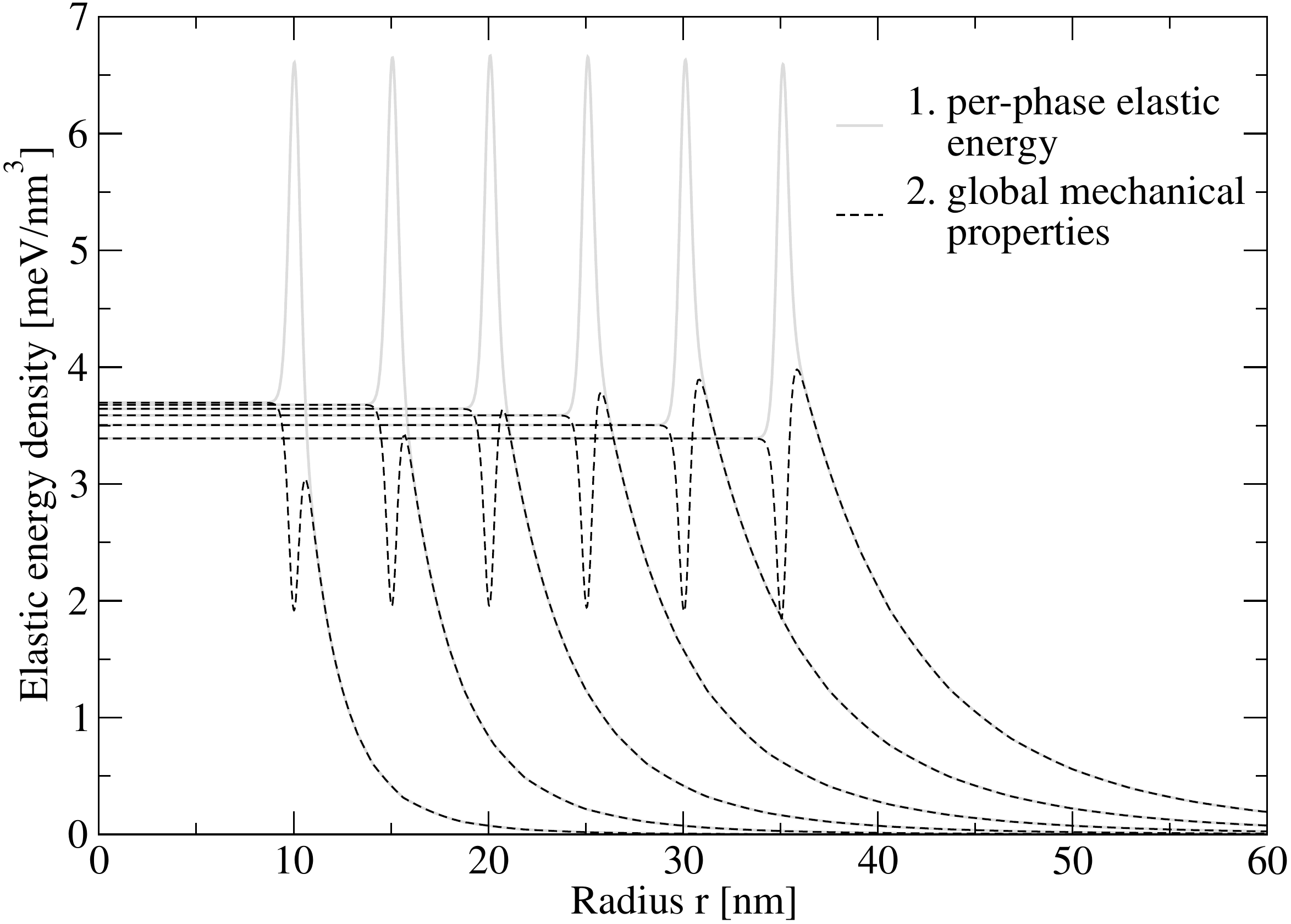}
  \caption{\label{fig:elastic} Radial elastic energy distribution of a set of six spherical particles with radii ranging from 10nm to 35nm embedded in a uniform matrix with mismatching eigenstrain. The two approaches of computing per-phase mechanical properties vs.\ global mechanical properties lead to energetic differences in the interface region of the particle.}
\end{figure}

The two approaches result in different elastic energy densities in interfacial regions. This is demonstrated in Fig.~\ref{fig:elastic}, which shows the elastic energy density of six spherical particles with varying radii as a function of distance from the particle center. The particle phase has a 5\% eigenstrain. Each phase has a parabolic free energy with an equilibrium solute concentration of 1 in the precipitate phase and 0 in the matrix phase. The multiphase model is assembled using~\eqref{eq:wbm}, with $W=4$, $h(\eta)=3\eta^2-2\eta^3$, and $g(\eta)=\eta^2(1-\eta)^2$. A single order parameter is used to model the two-phase system. The stiffness tensor is symmetric and isotropic with the non-zero entries being 1 eV/nm$^3$.

Note that the VTS per-phase elastic energy simulations result in a pronounced elastic energy excess at the interface. In both cases, the elastic strain state varies smoothly across the interface. In the Khachaturyan scheme, the effective eigenstrain also varies smoothly across the interface. In the VTS scheme each phase has a fixed eigenstrain value, and in the interface region the matrix is strongly stressed for both phases. This effect, and its consequences for simulating systems with strong elastic anisotropies, will be explored in a future publication.

\begin{figure}
  \centering
  \epsfig{width=0.9\linewidth,file=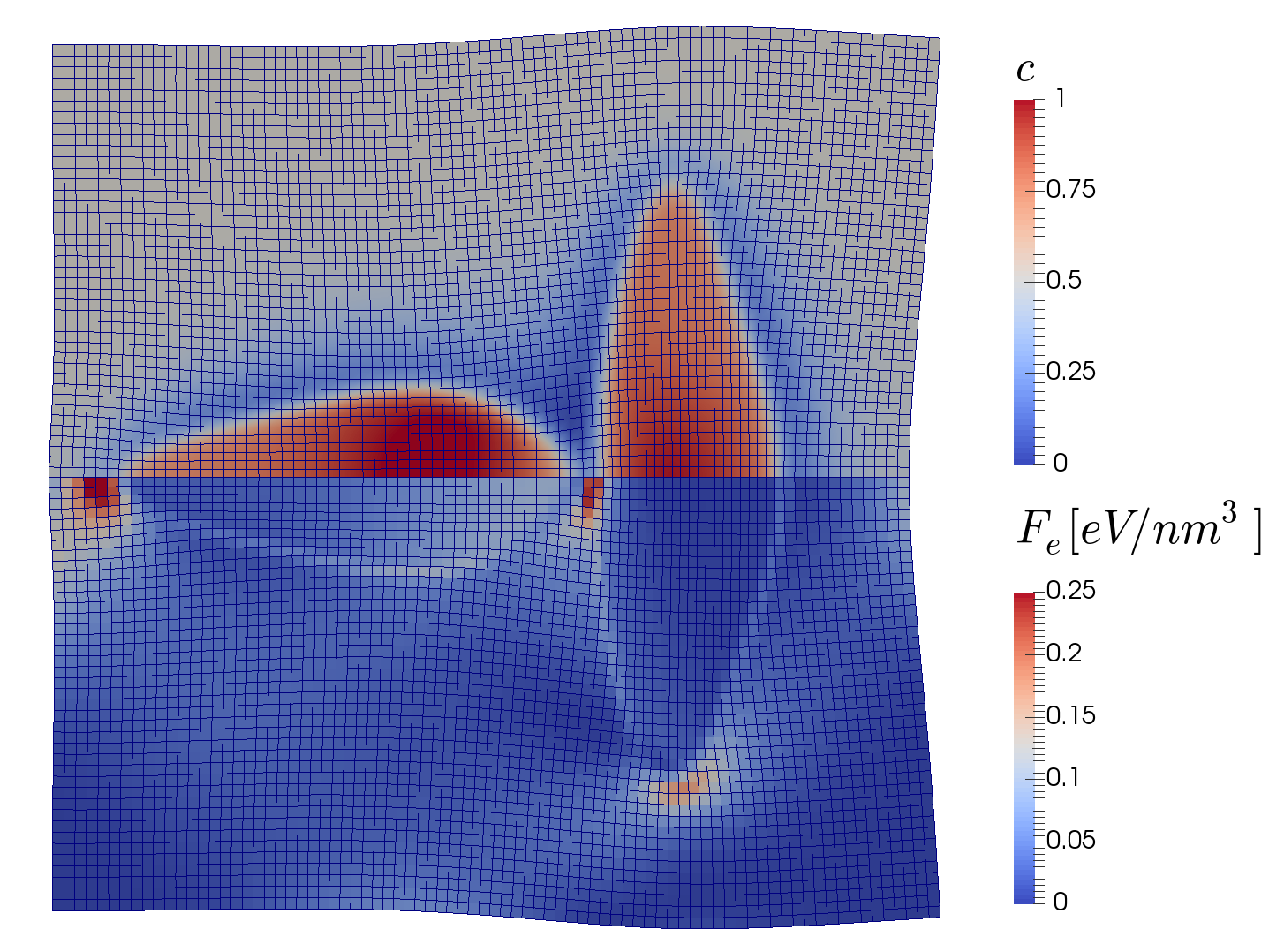}
  \caption{\label{fig:threephase} Three-phase precipitation problem with phase-field/mechanics coupling and anisotropic eigenstrains. The simulation cell is 40\,nm $\times$ 40\,nm. Displacements are exaggerated by a factor of 5.}
\end{figure}

As a further demonstration of the mechanics coupling capability, we consider an immiscible three-phase system consisting of a matrix phase and two precipitate phases with anisotropic eigenstrains simulated with the second (global mechanical properties) approach. Both particles have 5\% lattice contraction along their minor axis direction. The elastic properties of precipitates and matrix are set to a bulk modulus of 20 GPa and a shear modulus of 7 GPa. No-flux boundary conditions are applied for the phase-field variables and the null space for the displacement variables was eliminated by pinning select nodes. Fig.~\ref{fig:threephase} shows the simulation state after the precipitate growth has progressed. The mesh displacement is plotted with an amplification factor of 5. The bottom half of the plot shows the local strain energy density in eV/nm$^3$. The long range stress field in the minor axis direction enforces the lenticular shape of the particles. Both precipitate phases have a simple harmonic free energy with a minimum at $c=0.9$, while the matrix has its chemical free energy minimum at $c = 0$.
As an initial condition, two spherical nuclei with $c=0.9$  and a radius of 2\,nm were inserted in a super saturated matrix with $c=0.5$ to provide solute for particle growth.

\section{Conclusions}
In this work, we have summarized a novel capability for the rapid development of
multiphase-field models using automatic symbolic differentiation in the open source MOOSE framework. A modular free energy based approach allows researchers to focus on material model development without the need to touch the underlying numerical details of the coupled partial differential equation system solves. Encapsulating free energies together with their derivatives, which are needed by the phase field evolution equations, allows them to be recombined like building blocks at runtime to set up simulation scenarios. Both the automatic symbolic differentiation capability and the free energy based approach to the solution of phase-field models have been available in MOOSE for some time.

Symbolic differentiation enhances developer productivity in three important ways: it lowers the bar of entry required to investigate new/experimental phase-field models, it reduces the amount of time computational scientists must spend developing code, allowing that time to instead be spent on analysis, and it prevents an extremely common class of errors, i.e.\ incorrect Jacobians, from negatively impacting the efficiency and accuracy of results. We have shown that the performance of the automatically generated symbolic derivatives is at least on par with carefully handcrafted code when using the provided just-in-time compilation capability.

Leveraging the modular free energy system, we have implemented a set of multicomponent, multiphase models such as WBM and KKS that allow the user to combine arbitrary single phase free energies into multiphase free energies.
Tight coupling to linear elasticity is enabled though free energy modules that
provide the strain energy and its derivatives to the modular free
energy system. The various free energy contributions are combined at runtime, alleviating the need to modify and recompile code.
Together, these features offer increased flexibility when implementing multiphase models and solution methods, and when coupling to mechanics.


\section*{Acknowledgements}
This work was funded by the Department of Energy Nuclear Energy Advanced Modeling and Simulation program and the Light Water Reactor Sustainability program.  This manuscript has been authored by Battelle Energy Alliance, LLC under Contract No.~DE-AC07-05ID14517 with the US Department of Energy.
The publisher, by accepting the article for publication, acknowledges that the United States Government retains a nonexclusive, paid-up, irrevocable, world-wide license to publish or reproduce the published form of this manuscript, or allow others to do so, for United States Government purposes.

\section*{References}

\bibliography{ref}

\begin{thebibliography}{10}
\expandafter\ifx\csname url\endcsname\relax
  \def\url#1{\texttt{#1}}\fi
\expandafter\ifx\csname urlprefix\endcsname\relax\def\urlprefix{URL }\fi
\expandafter\ifx\csname href\endcsname\relax
  \def\href#1#2{#2} \def\path#1{#1}\fi

\bibitem{chen_phase-field_2002}
L.-Q. Chen, Phase-field models for microstructure evolution, Annual Review of
  Materials Research 32~(1) (2002) 113--140.
\newblock \href {http://dx.doi.org/10.1146/annurev.matsci.32.112001.132041}
  {\path{doi:10.1146/annurev.matsci.32.112001.132041}}.

\bibitem{moelans2008introduction}
N.~Moelans, B.~Blanpain, P.~Wollants, An introduction to phase-field modeling
  of microstructure evolution, CALPHAD 32~(2) (2008) 268--294.
\newblock \href {http://dx.doi.org/10.1016/j.calphad.2007.11.003}
  {\path{doi:10.1016/j.calphad.2007.11.003}}.

\bibitem{warren1995prediction}
J.~A. Warren, W.~J. Boettinger, Prediction of dendritic growth and
  microsegregation patterns in a binary alloy using the phase-field method,
  Acta Metallurgica et Materialia 43~(2) (1995) 689--703.
\newblock \href {http://dx.doi.org/10.1016/0956-7151(94)00285-P}
  {\path{doi:10.1016/0956-7151(94)00285-P}}.

\bibitem{karma1996phase}
A.~Karma, W.-J. Rappel, Phase-field method for computationally efficient
  modeling of solidification with arbitrary interface kinetics, Physical Review
  E 53~(4) (1996) R3017 (4 pages).
\newblock \href {http://dx.doi.org/10.1103/PhysRevE.53.R3017}
  {\path{doi:10.1103/PhysRevE.53.R3017}}.

\bibitem{wheeler1992phase}
A.~A. Wheeler, W.~J. Boettinger, G.~B. McFadden, Phase-field model for
  isothermal phase transitions in binary alloys, Physical Review A 45~(10)
  (1992) 7424--7439.
\newblock \href {http://dx.doi.org/10.1103/PhysRevA.45.7424}
  {\path{doi:10.1103/PhysRevA.45.7424}}.

\bibitem{kim_phase-field_1999}
S.~G. Kim, W.~T. Kim, T.~Suzuki, Phase-field model for binary alloys, Physical
  Review E 60~(6) (1999) 7186--7197.
\newblock \href {http://dx.doi.org/10.1103/PhysRevE.60.7186}
  {\path{doi:10.1103/PhysRevE.60.7186}}.

\bibitem{fan1997diffusion}
D.~Fan, L.-Q. Chen, Diffusion-controlled grain growth in two-phase solids, Acta
  Materialia 45~(8) (1997) 3297--3310.
\newblock \href {http://dx.doi.org/10.1016/S1359-6454(97)00022-0}
  {\path{doi:10.1016/S1359-6454(97)00022-0}}.

\bibitem{moelans2008quantitative}
N.~Moelans, B.~Blanpain, P.~Wollants, Quantitative analysis of grain boundary
  properties in a generalized phase field model for grain growth in anisotropic
  systems, Physical Review B 78~(2) (2008) 024113 (23 pages).
\newblock \href {http://dx.doi.org/10.1103/PhysRevB.78.024113}
  {\path{doi:10.1103/PhysRevB.78.024113}}.

\bibitem{wheeler1993computation}
A.~Wheeler, B.~Murray, R.~Schaefer, Computation of dendrites using a phase
  field model, Physica D: Nonlinear Phenomena 66~(1) (1993) 243--262.
\newblock \href {http://dx.doi.org/10.1016/0167-2789(93)90242-S}
  {\path{doi:10.1016/0167-2789(93)90242-S}}.

\bibitem{guyer2009}
J.~E. Guyer, D.~Wheeler, J.~A. Warren, {FiPy}: Partial differential equations
  with {P}ython, Computing in Science \& Engineering 11~(3) (2009) 6--15.
\newblock \href {http://dx.doi.org/10.1109/MCSE.2009.52}
  {\path{doi:10.1109/MCSE.2009.52}}.

\bibitem{takaki2005phase}
T.~Takaki, T.~Fukuoka, Y.~Tomita, Phase-field simulation during directional
  solidification of a binary alloy using adaptive finite element method,
  Journal of Crystal Growth 283~(1) (2005) 263--278.

\bibitem{tonks_object-oriented_2012}
M.~R. Tonks, D.~Gaston, P.~C. Millett, D.~Andrs, P.~Talbot, An object-oriented
  finite element framework for multiphysics phase field simulations,
  Computational Materials Science 51~(1) (2012) 20--29.
\newblock \href {http://dx.doi.org/10.1016/j.commatsci.2011.07.028}
  {\path{doi:10.1016/j.commatsci.2011.07.028}}.

\bibitem{chen1998applications}
L.-Q. Chen, J.~Shen, Applications of semi-implicit fourier-spectral method to
  phase field equations, Computer Physics Communications 108~(2) (1998)
  147--158.
\newblock \href {http://dx.doi.org/10.1016/S0010-4655(97)00115-X}
  {\path{doi:10.1016/S0010-4655(97)00115-X}}.

\bibitem{Puchala2016}
B.~Puchala, G.~Tarcea, E.~A. Marquis, M.~Hedstrom, H.~V. Jagadish, J.~E.
  Allison, The {M}aterials {C}ommons: A collaboration platform and information
  repository for the global materials community, JOM 68~(8) (2016) 2035--2044.
\newblock \href {http://dx.doi.org/10.1007/s11837-016-1998-7}
  {\path{doi:10.1007/s11837-016-1998-7}}.

\bibitem{logg2012automated}
A.~Logg, K.-A. Mardal, G.~Wells, Automated solution of differential equations
  by the finite element method: The FEniCS book, Vol.~84, Springer Science \&
  Business Media, 2012.

\bibitem{mmsp}
T.~Keller, et~al., The mesoscale microstructure simulation project,
  \url{https://github.com/mesoscale/mmsp} (2017).

\bibitem{AllenCahn}
S.~M. Allen, J.~W. Cahn, Ground state structures in ordered binary alloys with
  second neighbor interactions, Acta Metallurgica 20~(3) (1972) 423--433.
\newblock \href {http://dx.doi.org/10.1016/0001-6160(72)90037-5}
  {\path{doi:10.1016/0001-6160(72)90037-5}}.

\bibitem{CahnHilliard}
J.~W. {Cahn}, J.~E. {Hilliard}, {Free Energy of a Nonuniform System. I.
  Interfacial Free Energy}, The Journal of Chemical Physics 28~(2) (1958)
  258--267.
\newblock \href {http://dx.doi.org/10.1063/1.1744102}
  {\path{doi:10.1063/1.1744102}}.

\bibitem{Gaston_2015}
D.~R. Gaston, C.~J. Permann, J.~W. Peterson, A.~E. Slaughter, D.~Andr\v{s},
  Y.~Wang, M.~P. Short, D.~M. Perez, M.~R. Tonks, J.~Ortensi, R.~C. Martineau,
  Physics-based multiscale coupling for full core nuclear reactor simulation,
  Annals of Nuclear Energy, Special Issue on Multi-Physics Modelling of LWR
  Static and Transient Behaviour 84 (2015) 45--54.
\newblock \href {http://dx.doi.org/10.1016/j.anucene.2014.09.060}
  {\path{doi:10.1016/j.anucene.2014.09.060}}.

\bibitem{Brown_1986}
P.~N. Brown, A.~C. Hindmarsh, {Matrix-free methods for stiff systems of ODEs},
  SIAM Journal on Numerical Analysis 23~(3) (1986) 610--638.
\newblock \href {http://dx.doi.org/10.1137/0723039}
  {\path{doi:10.1137/0723039}}.

\bibitem{knoll:2004}
D.~A. Knoll, D.~E. Keyes, Jacobian-free {N}ewton-{K}rylov methods: A survey of
  approaches and applications, Journal of Computational Physics 193~(2) (2004)
  357--397.
\newblock \href {http://dx.doi.org/10.1016/j.jcp.2003.08.010}
  {\path{doi:10.1016/j.jcp.2003.08.010}}.

\bibitem{libMeshPaper}
B.~S. Kirk, J.~W. Peterson, R.~H. Stogner, G.~F. Carey, {\texttt{libMesh}: A
  C++ Library for Parallel Adaptive Mesh Refinement/Coarsening Simulations},
  Engineering with Computers 22~(3--4) (2006) 237--254.
\newblock \href {http://dx.doi.org/10.1007/s00366-006-0049-3}
  {\path{doi:10.1007/s00366-006-0049-3}}.

\bibitem{petsc-user-ref}
S.~Balay, S.~Abhyankar, M.~F. Adams, J.~Brown, P.~Brune, K.~Buschelman,
  L.~Dalcin, V.~Eijkhout, W.~D. Gropp, D.~Kaushik, M.~G. Knepley, L.~C.
  McInnes, K.~Rupp, B.~F. Smith, S.~Zampini, H.~Zhang, H.~Zhang,
  \href{http://www.mcs.anl.gov/petsc}{{PETS}c users manual}, Tech. Rep.
  ANL-95/11 - Revision 3.7, Argonne National Laboratory (2016).
\newline\urlprefix\url{http://www.mcs.anl.gov/petsc}

\bibitem{zhang_quantitative_2013}
L.~Zhang, M.~R. Tonks, D.~Gaston, J.~W. Peterson, D.~Andrs, P.~C. Millett,
  B.~S. Biner, A quantitative comparison between and elements for solving the
  cahn-hilliard equation, Journal of Computational Physics 236 (2013) 74--80.
\newblock \href {http://dx.doi.org/10.1016/j.jcp.2012.12.001}
  {\path{doi:10.1016/j.jcp.2012.12.001}}.

\bibitem{fparser-web-page}
J.~Nieminen, J.~Yliluoma,
  \href{http://warp.povusers.org/FunctionParser}{{F}unction {Parser} {W}eb
  page} (2011).
\newline\urlprefix\url{http://warp.povusers.org/FunctionParser}

\bibitem{TOLSMA1998475}
J.~E. Tolsma, P.~I. Barton, On computational differentiation, Computers \&
  Chemical Engineering 22~(4--5) (1998) 475--490.
\newblock \href {http://dx.doi.org/10.1016/S0098-1354(97)00264-0}
  {\path{doi:10.1016/S0098-1354(97)00264-0}}.

\bibitem{Kedem1980}
G.~Kedem, Automatic differentiation of computer programs, ACM Transactions on
  Mathematical Software (TOMS) 6~(2) (1980) 150--165.
\newblock \href {http://dx.doi.org/10.1145/355887.355890}
  {\path{doi:10.1145/355887.355890}}.

\bibitem{dlopen-page}
\href{http://www.opengroup.org/susv3xsh/dlopen.html}{{The Open Group Base
  Specifications Issue 6 -- dlopen}} (2004).
\newline\urlprefix\url{http://www.opengroup.org/susv3xsh/dlopen.html}

\bibitem{sha1}
Q.~H. Dang, {Secure Hash Standard (SHS)}, Tech. Rep. FIPS-180-4, Information
  Technology Laboratory, National Institute of Standards and Technology, our
  implementation is from: \url{http://www.tamale.net/sha1} (Mar. 2012).
\newblock \href {http://dx.doi.org/10.6028/NIST.FIPS.180-4}
  {\path{doi:10.6028/NIST.FIPS.180-4}}.

\bibitem{Lattner2004}
C.~Lattner, V.~Adve,
  \href{http://dl.acm.org/citation.cfm?id=977395.977673}{{LLVM}: {A}
  compilation framework for lifelong program analysis \& transformation}, in:
  Proceedings of the International Symposium on Code Generation and
  Optimization: Feedback-directed and Runtime Optimization (CGO), 2004, pp.
  75--87.
\newline\urlprefix\url{http://dl.acm.org/citation.cfm?id=977395.977673}

\bibitem{Clang}
C.~Lattner, et~al., \href{http://clang.llvm.org/}{clang: A {C} language family
  frontend for {LLVM}} (2014).
\newline\urlprefix\url{http://clang.llvm.org/}

\bibitem{GCC}
R.~Stallman, et~al., \href{https://gcc.gnu.org/}{{GCC}, the {GNU} compiler
  collection} (2014).
\newline\urlprefix\url{https://gcc.gnu.org/}

\bibitem{Hildebrand}
J.~H. Hildebrand, The term `regular solution', Nature 168 (1951) 868.
\newblock \href {http://dx.doi.org/10.1038/168868a0}
  {\path{doi:10.1038/168868a0}}.

\bibitem{Jokisaari2015334}
A.~Jokisaari, K.~Thornton, General method for incorporating {CALPHAD} free
  energies of mixing into phase field models: Application to the
  $\alpha$-zirconium/$\delta$-hydride system, CALPHAD 51 (2015) 334--343.
\newblock \href {http://dx.doi.org/10.1016/j.calphad.2015.10.011}
  {\path{doi:10.1016/j.calphad.2015.10.011}}.

\bibitem{schwen_analytic_2013}
D.~Schwen, E.~Martinez, A.~Caro, On the analytic calculation of critical size
  for alpha prime precipitation in {FeCr}, Journal of Nuclear Materials
  439~(1-3) (2013) 180--184.
\newblock \href {http://dx.doi.org/10.1016/j.jnucmat.2013.03.057}
  {\path{doi:10.1016/j.jnucmat.2013.03.057}}.

\bibitem{li_phase-field_2013}
Y.~Li, S.~Hu, R.~Montgomery, F.~Gao, X.~Sun, Phase-field simulations of
  intragranular fission gas bubble evolution in {UO}$_2$ under post-irradiation
  thermal annealing, Nuclear Instruments and Methods in Physics Research
  Section B: Beam Interactions with Materials and Atoms 303 (2013) 62--67.
\newblock \href {http://dx.doi.org/10.1016/j.nimb.2012.11.028}
  {\path{doi:10.1016/j.nimb.2012.11.028}}.

\bibitem{Ohno2010}
M.~Ohno, K.~Matsuura, Quantitative phase-field modeling for two-phase
  solidification process involving diffusion in the solid, Acta Materialia
  58~(17) (2010) 5749--5758.
\newblock \href {http://dx.doi.org/10.1016/j.actamat.2010.06.050}
  {\path{doi:10.1016/j.actamat.2010.06.050}}.

\bibitem{Folch2005}
R.~Folch, M.~Plapp, Quantitative phase-field modeling of two-phase growth,
  Physical Review E 72~(1) (2005) 011602 (27 pages).
\newblock \href {http://dx.doi.org/10.1103/PhysRevE.72.011602}
  {\path{doi:10.1103/PhysRevE.72.011602}}.

\bibitem{Toth2015}
G.~I. T\'{o}th, T.~Pusztai, L.~Gr\'{a}n\'{a}sy, Consistent multiphase-field
  theory for interface driven multidomain dynamics, Physical Review B 92~(18)
  (2015) 184105 (19 pages).
\newblock \href {http://dx.doi.org/10.1103/PhysRevB.92.184105}
  {\path{doi:10.1103/PhysRevB.92.184105}}.

\bibitem{GyoonKim2004135}
S.~G. Kim, W.~T. Kim, T.~Suzuki, M.~Ode, Phase-field modeling of eutectic
  solidification, Journal of Crystal Growth 261~(1) (2004) 135--158.
\newblock \href {http://dx.doi.org/10.1016/j.jcrysgro.2003.08.078}
  {\path{doi:10.1016/j.jcrysgro.2003.08.078}}.

\bibitem{Ammar2009}
K.~Ammar, B.~Appolaire, G.~Cailletaud, S.~Forest, Combining phase field
  approach and homogenization methods for modelling phase transformation in
  elastoplastic media, European Journal of Computational Mechanics 18~(5-6)
  (2009) 485--523.
\newblock \href {http://dx.doi.org/10.3166/ejcm.18.485-523}
  {\path{doi:10.3166/ejcm.18.485-523}}.

\bibitem{Khachaturyanbook}
A.~G. Khachaturyan, Theory of Structural Transformations in Solids, Wiley, New
  York, 1983.

\bibitem{Vaithyanathan2002}
V.~Vaithyanathan, C.~Wolverton, L.-Q. Chen, Multiscale modeling of precipitate
  microstructure evolution, Physical Review Letters 88~(12) (2002) 125503 (4
  pages).
\newblock \href {http://dx.doi.org/10.1103/PhysRevLett.88.125503}
  {\path{doi:10.1103/PhysRevLett.88.125503}}.

\bibitem{Vaithyanathan2004}
V.~Vaithyanathan, C.~Wolverton, L.-Q. Chen, Multiscale modeling of $\theta'$
  precipitation in {A}l-{C}u binary alloys, Acta Materialia 52~(10) (2004)
  2973--2987.
\newblock \href {http://dx.doi.org/10.1016/j.actamat.2004.03.001}
  {\path{doi:10.1016/j.actamat.2004.03.001}}.

\end{thebibliography}

\end{document}